\documentclass{llncs} 

\usepackage[colorlinks]{hyperref}
\usepackage{breakurl}
\usepackage{listings}
\usepackage{graphicx}
\usepackage{amstext}
\usepackage{bbm}
\usepackage{amsmath}
\usepackage{amsfonts}
\usepackage{amssymb}
\usepackage{xspace}
\usepackage{multirow}

\usepackage{color}

\newcommand{\apriori}{\emph{a priori}\xspace}

\newcommand{\vect}[1]{\ensuremath{\mathbf{#1}}}
\newcommand{\abs}[1]{\ensuremath{\left\lvert#1\right\rvert}}

\begin{document}

\title{e{X}pose: A Character-Level Convolutional Neural Network with Embeddings For Detecting Malicious URLs, File Paths and Registry Keys}

\author{Joshua Saxe\inst{1} and Konstantin Berlin\inst{2}}
\institute{
Invincea Inc.\\
\email{josh.saxe@invincea.com}\and
Invincea Inc.\\
\email{kberlin@invincea.com}
}

\maketitle

\begin{abstract}

For years security machine learning research has promised to obviate the need for signature based detection by automatically learning to detect indicators of attack.  Unfortunately, this vision hasn't come to fruition: in fact, developing and maintaining today's security machine learning systems can require  engineering resources that are comparable to that of signature-based detection systems, due in part to the need to develop and continuously tune the ``features'' these machine learning systems look at as attacks evolve.  Deep learning, a subfield of machine learning, promises to change this by operating on raw input signals and automating the process of feature design and extraction.  In this paper we propose the eXpose neural network, which uses a deep learning approach we have developed to take generic, raw short character strings as input (a common case for security inputs, which include artifacts like potentially malicious URLs, file paths, named pipes, named mutexes, and registry keys), and learns to simultaneously extract features and classify using character-level embeddings and convolutional neural network.  In addition to completely automating the feature design and extraction process, eXpose outperforms manual feature extraction based baselines on all of the intrusion detection problems we tested it on, yielding a 5\%-10\% detection rate gain at $0.1$\% false positive rate compared to these baselines.

\end{abstract}

\section{Introduction}

While for over a decade researchers have proposed systems that apply machine learning methods to computer security detection problems, this research has gained only limited prevalence in real-world security systems, in part, we believe, because machine learning systems require significant expert effort to develop and maintain.

For example, development of machine learning based security detection systems requires an in-depth exploration of the feature representation of a given security artifact type (e.g. Windows PE binaries, URLs, or behavioral traces), and an exploration of what machine learning detection approaches yield the best accuracy given those representations.  As cyber-attacks evolve, machine learning feature representations must be updated to keep pace with the latest cyber threats.  The calculation of many computer security products companies is often that signature based systems are thus a less risky investment.

While many technical problems stand in the way of effective deployment of machine learning systems (e.g. the collection of large volumes of labeled training data, the problem of evaluating these systems when attacker behavior is constantly changing, and the problem of deploying complex models of low-resource endpoints), one way to reduce the cost of creating and maintaining machine learning approaches is to move beyond manual feature engineering, given that feature engineering is often recognized as the most time consuming aspect of machine learning system development.  Deep learning, a subfield of machine learning that utilizes neural networks operating directly on raw inputs, promises to allow us to do this.

In line with this vision, we present eXpose, a deep learning approach to a number of security detection problems, that directly works on raw inputs to detect maliciousness. Specifically, eXpose takes generic short character strings as its input and learns to detect whether they are indicators of malicious behavior based on their lexical semantics.  In this paper, we demonstrate eXpose's ability to detect malicious URLs, malicious file paths, and malicious registry keys.

To make our research objectives clear, below are examples of these data, starting with malicious URLs (we've substituted URL forward slashes for backslashes to avoid accidental clicks):
\begin{verbatim}
http:\\0fx8o.841240.cc\201610\18\content_23312\svchost.exe
http:\\31.14.136.202\secure.apple.id.login\Apple\login.php
http:\\1stopmoney.com\paypal-login-secure\websc.php
\end{verbatim}
Next, a few examples of malicious file paths:
\begin{verbatim}
C:\Temp\702D97503A79B0EC69\JUEGOS/Call of Duty 4+Keygen
C:\Temp\svchost.vbs
C:\DOCUME~1\BASANT~1\LOCALS~1\Temp\WzEC.tmp\fax.doc.exe
\end{verbatim}
Finally, a few examples of malicious registry keys:
\begin{verbatim}
HKCU\Software\Microsoft\Windows\CurrentVersion\Run Alpha Antivirus
HKCR\Applications\WEBCAM HACKER 1.0.0.4.EXE
HKCR\AppID\bccicabecccag.exe
\end{verbatim}

All of these examples appear malicious, or at least suspicious, to the expert eye, leading us to hypothesize that a machine learning system could also infer their maliciousness. It might even be possible to exceed human expert's ability to guess whether these artifacts are malicious, by learning to recognize generalized deceptive patterns observed over tens of millions of malicious artifacts.  And indeed, on all of the intrusion detection problems we tested, eXpose outperformed manual feature extraction based machine learning baselines, yielding a 5\%-10\% higher detection rate at deployment relevant false positive rates.  Our research demonstrates the potential deep learning methods hold solving hard security detection problems.

The rest of this paper is structured as follows. In Section \ref{sec:prevwork} we describe related work. In Section \ref{sec:method} we motivate and describe our eXpose approach, including an exact and reproducible description of our system. Section \ref{sec:results} describes our evaluation methodology and results.  Finally, in section \ref{sec:conclusions} we sum up the paper and discuss directions for future work.

\section{Previous Work}
\label{sec:prevwork}

\subsection{Related work in computer security}

We designed eXpose as a fairly generic detection tool that simultaneously addresses a number of cybersecurity problems. This is somewhat different from previous work, which tends to focus on individual security detection problems, such as identifying malicious URLs or malicious host-based behavior individually.  In this section we place our work in conversation with current cybersecurity literature and describe its relationship to the broader deep learning literature.

A number of previous works in machine learned based behavioral detection of malware is related to automatic classification of individual file paths or registry keys.  In general, previous behavioral malware detection methods have focused on making detections on the basis of sequences of observed process or operating system-level events.  For example, \cite{berlin2015malicious} proposes a logistic regression-based method for detecting malware infections based on n-grams of audit log event observations.  Relatedly, \cite{apap2002detecting} proposes to use an anomaly detection approach on sequences of registry accesses to infer whether a host has been compromised.  \cite{firdausi2010analysis} surveys a wide variety of behavioral malware detection techniques, all of which perform manual feature engineering on collections of events to infer whether or not dynamically executed binaries are malicious or benign.

Unlike the work summarized above, which operates on groups of dynamic host-based observations to detect malware, eXpose operates on individual events, but rather than modeling individual host-based events using manually defined feature representations, eXpose learns representations of input strings (e.g. file paths and registry keys) as part of its overall process of learning to make accurate detections of malicious behavior.  We thus think of eXpose as providing a complementary and orthogonal detection capability relative to these  research efforts.

Unlike individual file and registry writes, identifying malicious URLs is a more studied problem in the security detection literature. Proposed malicious URL detection approaches have tended to either exclusively use URL strings as their input or utilize both URL strings and supplementary information like website registration services, website content, and network reputation \cite{khonji2013phishing}.  In contrast to work that uses both input URLs and auxiliary information to detect malicious URLs, our work relies solely on URL input strings, making it easier to deploy.

With respect to the detection mechanism used in previous URL detection work, the simplest proposed approaches have involved blacklists, which can be collected using manual labeling, user feedback, client honeypots, and other heuristics \cite{huang2014malicious}. While blacklists have a very low false positive rate, they are also very brittle and thus cannot generalize to previously unseen URL strings \cite{sheng2009empirical}. To address these limitations, statistical approaches, such as machine-learning or similarity based URL detection have been proposed \cite{khonji2013phishing}.  Unfortunately, manually discovering potentially useful features is time consuming and requires constant adaptation to evolving obfuscation techniques, which limits the achievable accuracy of the detectors.  In contrast to work that requires manual feature extraction from URLs to make detections, our work automates this feature extraction process.

\subsection{Machine Learning}

\subsubsection{Convolutional Neural Networks}

eXpose uses neural network convolutional kernels as part of its approach to automating feature engineering and extraction, and so in addition to computer security literature focused on detecting cyber attacks, our work is also related to the convolutional neural network and recurrent neural network literature in machine learning.

Convolutional Neural Networks (CNN) \cite{lecun1989backpropagation,lecun1995learning} have been applied to image recognition problems for a long time, but only fairly recently have they demonstrated breakthrough results in image recognition \cite{krizhevsky2012imagenet}. The advantage of CNNs over previous approaches is that they work directly on the raw pixel data, thus eliminating tedious and fairly limited hand designing of features. What makes CNNs particularly powerful for images is that they are able to efficiently exploit information locality by applying convolutional operations on a raw data using a set of different kernels. These kernels are learned jointly with the entire network, and thus are better able to adapt to learning objectives than hand designed designed kernels. Since the same kernel is applied to every pixel of the image, there is a tremendous reduction in parameters that need to be learned, as compared to a fully dense neural network.

In addition to targeting image inputs, CNNs have also found rich applications within natural language processing, where they are typically applied to subsequences of words or patterns such that they perform pattern matching on sequential patterns within input texts.  For example, \cite{severyn2015unitn} proposes to combine a word embedding approach with a convolutional neural network to perform sentiment analysis on Twitter data.

Other authors propose machine learning approaches that operate on character level embeddings \cite{cicero2014deep,kim2014convolutional,zhang2015character}. The advantage of such approaches is that they do not require syntactical understanding of the language, such as word boundaries or punctuation. This work is closely to related to our own, since we also model text strings at the character level, embedding them in an embedding space and then extracting features using convolutions.  To our knowledge, our approach is the first computer security detector to take this approach or any approach that extracts features from raw inputs.

\subsubsection{Recurrent Neural Networks}

RNNs are commonly used to process sequential information, with LSTM based approaches being among the more popular \cite{hochreiter1997long}. While theoretically they are able to learn long terms dependencies in a sequences, RNNs are problematic to train due to the vanishing gradient problem as well as large computational cost because of the need to sequentially back and forward propagate information during training and prediction phases \cite{hochreiter2001gradient}. Recently CNN have been shown to be just (if not more) effective on sequential data oriented modeling problems, while allowing a significantly faster model training and prediction evaluation \cite{kalchbrenner2016neural}.

\subsubsection{Multi-task Learning}

Multi-task learning using neural networks is a somewhat related idea to our generalized CNN model. There, a set of network layers is shared between various related learning tasks, and an individualized set of final layers is used to make the final prediction on the specific task \cite{caruana1993multitask}. Sharing internal weights potentially allows the network to learn a richer representation of the data, which is especially useful when individual datasets are very small \cite{baxter1995learning}. Another advantage of multi-task learning is if a set of related classifiers needs to be deployed to an endpoint, weight-sharing provides a smaller deployment footprint.

Multi-task learning has also been suggested for malware detection by simultaneously trying to predict binary classification as well as malware family \cite{huang2016mtnet}. However, multi-task learning does not directly map to our set of detection problems, where the semantic meaning of the characters and substrings significantly changes between problems. Furthermore, our labeled dataset is of virtually unlimited size, making training on combined datasets less useful.

\section{Method}
\label{sec:method}

eXpose is built on the premise that applying a neural network directly to the raw input of short character strings provides better classification accuracy than previous approaches that rely on hand-designed features. In this section we describe how we architected our neural network to operate directly on raw character input, and the intuition behind our decisions. Our network was implemented in Python 2.7 using Keras v1.1 \cite{chollet2015keras}.

\subsection{Architecture}

\begin{figure*}[thpb]
      \centering
      \includegraphics[width=1.0\textwidth]{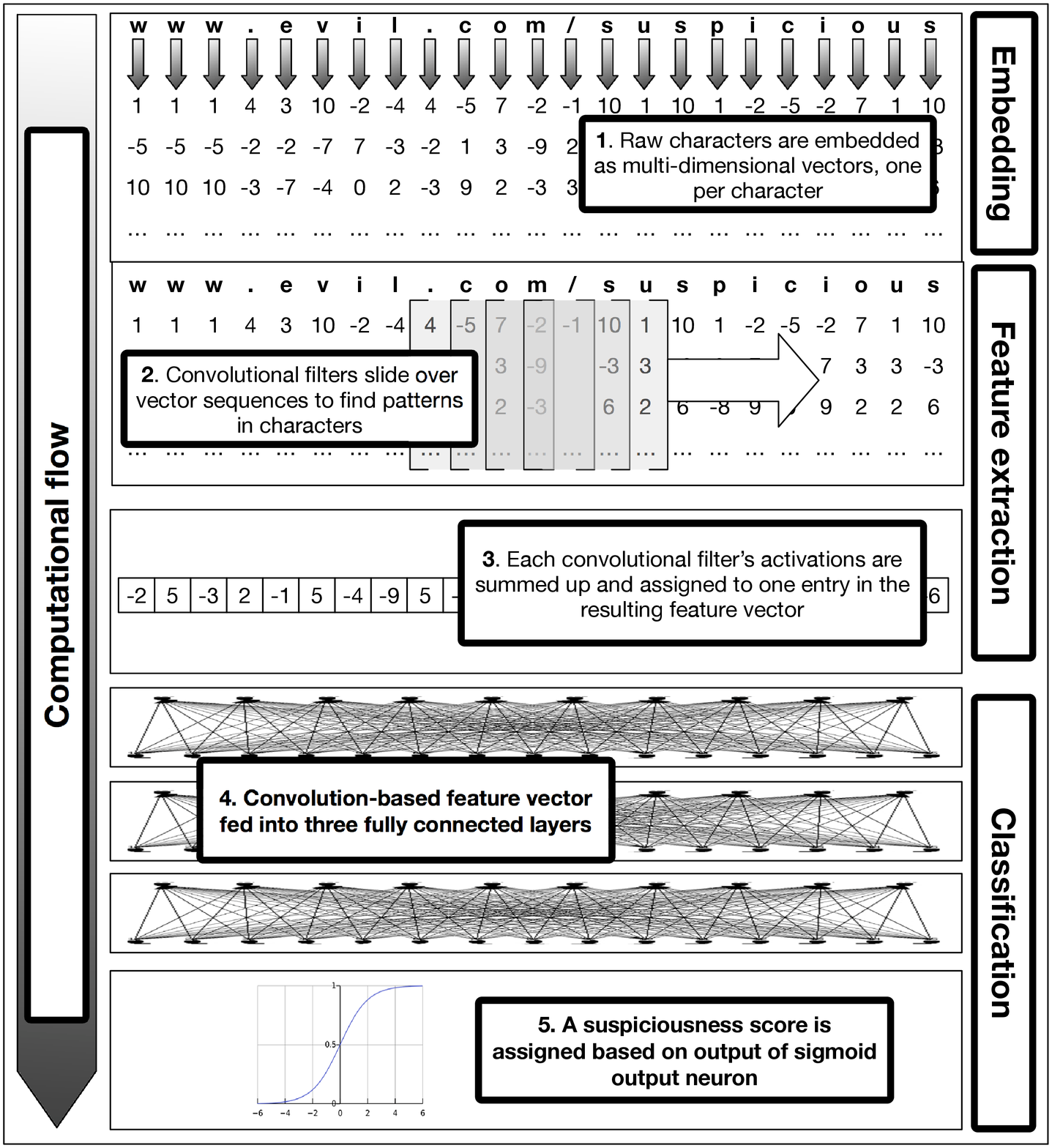}
      \caption{An intuitive overview of the neural network architecture of our model. }
      \label{fig:accessible_overview_eps}
\end{figure*}

Fig. \ref{fig:accessible_overview_eps} gives an intuitive overview of our approach, showing that our neural network is divided into three notional components \footnote{What we mean here is that our overall model is most easily understood as containing three separate components, each focused on a somewhat different task. We used this notional hierarchy when developing our networks architecture. It is important to note, however, that the entire model is simultaneously optimized, end-to-end, and thus all components are optimized for the singular classification task. We can think of the entire model as some complex classifier, or alternatively, a deep feature extractor, followed by a logistic regression.}: character embedding, feature detection, and a classifier.

The character embedding component embeds the alphabet of printable English-language characters into a multi-dimensional feature space, thus encoding an input string's sequence of raw characters as a two-dimensional tensor.

Using this tensor, the feature detection component detects important local sequence patterns within the full character sequence, and then aggregates this information into a fixed-length feature vector.

\begin{figure*}[thpb]
      \centering
      \includegraphics[width=1.0\textwidth]{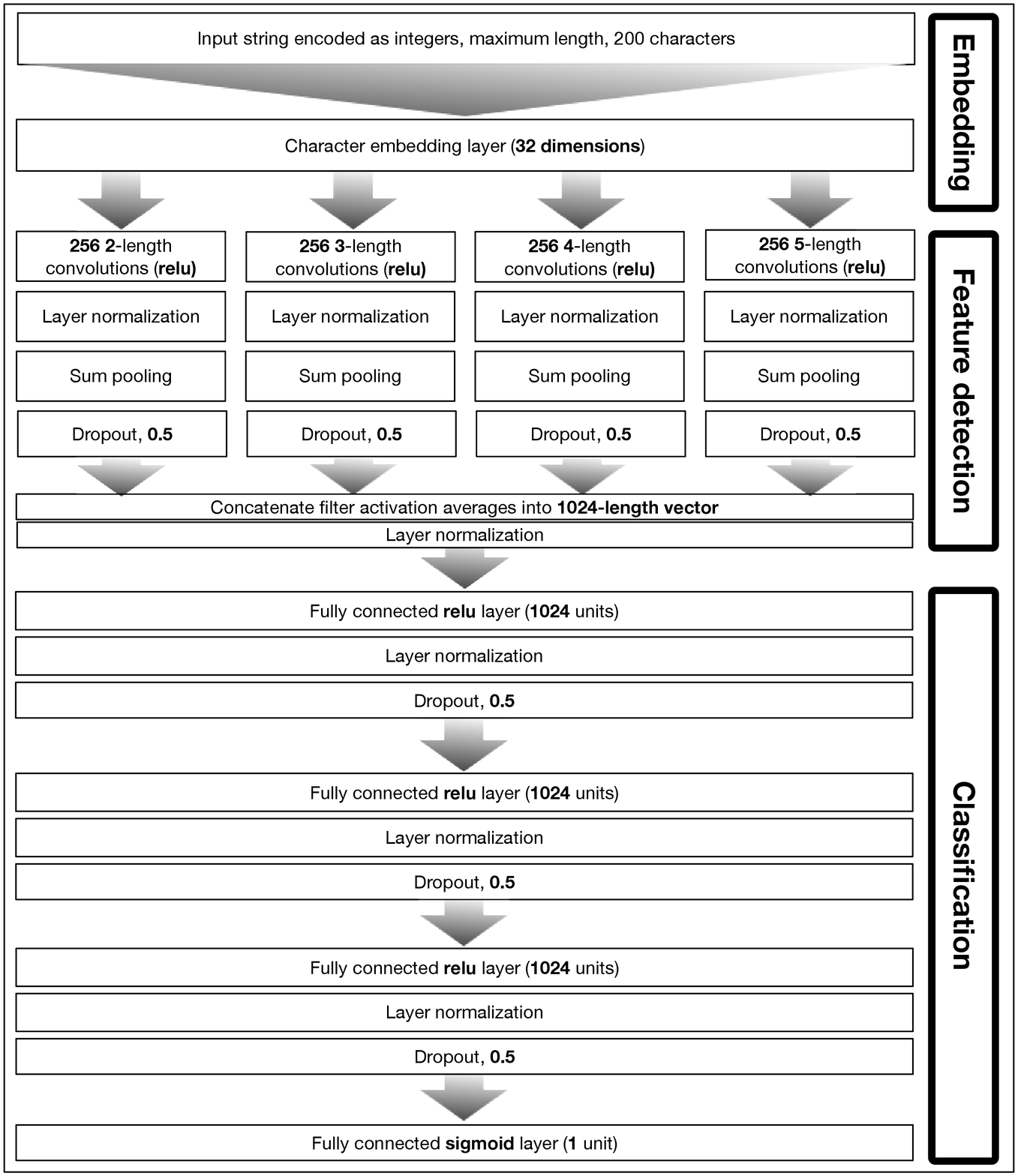}
      \caption{The neural network architecture of our CNN model. }
      \label{fig:inaccessible_overview_eps}
\end{figure*}

Finally, the classification component classifies the detected features using a dense neural network.  All of these components are optimized jointly using stochastic gradient descent.  Fig. \ref{fig:inaccessible_overview_eps} gives a formal diagram of our neural network architecture, which we also describe step by step in the text below.

\subsubsection{Character Embedding}

Our model starts its computational flow with the raw length $s$ sequence of characters and embeds them into an $s\times m$ floating point matrix. This operation is a simply dictionary lookup, where each character, irrespective of the characters that came before it or after it, is mapped to its corresponding vector, and then these vectors are concatenated into this matrix. The matrix's rows  represent the sequence of characters in the original string, and the matrix's columns represent the dimensions of the embedding space.

Embedding layer is optimized jointly with the rest of the model through back-propagation, optimizing the individual characters' embedding vectors to be more reflective of their semantic meaning, resulting in pairs of semantically similar characters being embedded closer to each other if they have similar attributes (e.g. they are both uppercase, both control control characters, etc.) \cite{kim2014convolutional}. This clustering of semantically similar characters makes it similar for the lower layers to identify semantically similar patterns in the string.

In our implementation we set $s=200$ and $m=32$.  For our URL-based experiments we use an input vocabulary of the 87 URL-valid characters, and for our file path and registry key experiments we use an input vocabulary of the 100 valid printable characters.  Any unicode we encounter in our experiments we wildcard with a lower-case `x', reserving more sophisticated handling of international characters for future work.  We set maximum string length on all artifacts to $200$ since this is around the $95$th percentile or greater of all strings in our experimental URL, file path, and registry datasets. If the string is shorter than $200$, we pad it with a special null symbol in the front. If the string is longer than $200$, we cutoff the beginning of the string. We empirically determined that $m=32$ provides a good tradeoff between accuracy and computational complexity. Note that $32$ is much smaller then the potential $87$-sized or $100$-sized one-hot-encoding of these same characters, which is a common way to represent categorical input in machine learning models.

We visually demonstrate that our trained model does indeed learn semantically related embedding in Fig. \ref{fig:results}, where we show a two-dimensional MDS projection of our learned embeddings. As the Fig. \ref{fig:results} shows, letters with similar semantics tend to cluster together, with upper case letters appearing near other upper case letters, lowercase letters near other lowercase letters, tilde, a highly important character in the URL case falling into its own cluster, etc.  The fact that our character vectors cluster in this way suggests that our embedding representation is capturing their semantic meaning.

\subsubsection{Feature Detection}

Once we embed our input into an $s \times m$ matrix, the next step is extracting and aggregating locally detected features. This is done in two stages: in the first stage we detect local features by applying multiple kernel convolutions, $\mathrm{Conv}(t, k, n)$, and in the second stage we aggregate the results across the entire sequence by summing the kernels' activations using $\mathrm{SumPool}$.  We define both $\mathrm{Conv}$ and $\mathrm{SumPool}$ formally in our appendix.  These steps are done separately for each $k \in \{2,3,4,5\}$, and we empirically set $t=256$. The four results for each $k$ tower are then concatenated together into a $1024$ length vector.

The $t$ filters in our CNN spans the entire length of the character embedding $m$, and can be thought of as “sliding” of convolution kernels (or masks) over the sequence of character embeddings. The motivation for using convolutions as our feature extraction component flows from similar approaches in natural language processing (NLP) \cite{zhang2015character}.  

Computing convolutions on raw character embedding matrix is conceptually similar to traditional bag-of-words approaches. The main conceptional difference is that rather than directly detecting n-grams, we allow for ``approximate'' matches on semantically similar substrings.

In this manner of thinking, each convolutional filter is responsible for detecting a distinct set of similar sequential patterns, and by summing up its activations over a text string, we obtain the degree to which these patterns occurs in the full string, similar to a bag-of-words aggregating all the n-grams.  Just like the n-gram approach, our approach is robust to insertions and deletions within the character-string, as subsequences can occur anywhere in the string and still be detected by the convolution.

\subsubsection*{Normalization and Regularization}

To speed up model training and prevent overfitting, we use layer-wise $\mathrm{BatchNorm}$ and $\mathrm{Dropout}(0.5)$ ($0.2$ for registry keys) between layers (see Fig. \ref{fig:inaccessible_overview_eps} for details).  We define both $\mathrm{BatchNorm}$ and $\mathrm{Dropout}$ in our appendix below.  We found that layer-wise normalization gave better results than the more popular batch normalization \cite{ioffe2015batch}, and that putting the normalization after the activation gave equivalent results to putting it before each unit’s activation function.  We also found that without regularization our model can easily overfit our training data, even when training on millions of samples.

\subsubsection{Classification}

Once we extract the features, we use a standard dense neural network to classify the string as malicious or benign.  The dense neural network has two layers, a $\mathrm{Dense}(l)$ unit, followed by the $\mathrm{DenseSigmoid}(l)$ layer with $l=1024$ units. We define both $\mathrm{Dense}$ and $\mathrm{DenseSigmoid}$ in our appendix below.  The dense layers learn a non-linear kernel given the convolution based features, and the sigmoid layer output provides the probability that the input string is malicious given the output of the final dense layer. We measure our detector's prediction loss using binary-cross entropy,
\begin{equation}
	\mathcal{L}(\vect{\hat{y}}, \vect{y}) = -\frac{1}{N}\sum_i^N \left[ y_i \log \hat{y}_i + (1-y_i) \log (1-\hat{y}_i) \right]
\label{eq:loss}
\end{equation}
where $\vect{\hat{y}}$ is our model's prediction probability vector for all the URL samples and $\vect{y}$ is the vector of true label ($0$ for benign, and $1$ for malicious). We use Adam \cite{kingma2014adam} method to minimize eq. \eqref{eq:loss}.

Typically, it is much easier to collect benign than malicious data, resulting in highly imbalanced dataset. Rather than simply reweighing individual values to equalize the overall contribution of the benign and malware in eq. \eqref{eq:loss}, we adjusted the benign to malware ratio directly during batch streaming. We generate 256-sized batch by first randomly sampling the full dataset of benign samples and randomly selected 128 samples, and then repeating the same approach for 128 malware samples. This effectively created an even class balance between malicious and benign data in our training dataset, with a more diverse representation of each class than simple reweighting of individual samples. We count one epoch as having processed 4096 batches, and train for 100 epochs.

For our final solution we select the best overall model, determined by largest area under the ROC (AUC) for the time-split validation.

\subsection*{Alternatives Attempted}

During the development of our current architecture we have also tried several alternative architectures that did not yield better performance, and/or where computationally too expensive: 
\begin{itemize}
\item Replacing regular convolutions with a series of dilated convolutions \cite{yu2016multi,kalchbrenner2016neural}.
\item Stacking embeddings, convolutions, and a long short term memory recurrent layer \cite{sainath2015convolutional}.
\item Stacked convolutions to learn non-linear convolutional activations \cite{szegedy2015going}.
\item ResNet type stacking of convolutions \cite{he2016identity}
\end{itemize}

\section{Results}
\label{sec:results}

We evaluate our model against two baseline approaches, described below, on three different problems. The first problem involves identifying malicious URLs directly from the URL string. For this problem we downloaded 19067879 unique URLs, randomly sampled over a roughly two month period, from VirusTotal.

For the second and third problem, identifying malicious registry key and file paths, we extracted over 18 million Cuckoo sandbox runs, as recorded on VirusTotal, and utilized all the observed file and registry writes and creations. This gave us 5590614 unique file paths, and 1661716 registry key paths. We give a detailed breakdown of these data in Table \ref{tab:data}.

\begin{table*}[h]
\caption{Numbers of artifacts in our research datasets}
\label{tab:data}
\begin{center}
\begin{tabular}{|l|r|r|r|r|r|r|}
\hline
\multirow{2}{*}{Type} & \multicolumn{2}{|c|}{Training} & \multicolumn{2}{|c|}{Testing} \\
 & Benign & Malicious & Benign & Malicious \\
 \hline
URLs & 7211705 & 1496198 & 9718748 & 641228 \\
File Paths & 869836 & 3677404 & 359796 & 683578 \\
Registry Keys & 250819 & 1282292 & 43437 & 85168 \\
\hline
\end{tabular}
\end{center}
\end{table*}

\subsection{Labeling}

Training and evaluating eXpose's performance requirs assigning a binary label to every artifact in our experimental datasets indicating whether it is malicious or benign.  We label URL artifacts we used a voting approach, in which we assigned a label to the URL based on the score given by 59 anti-virus engines.  If 5 or more of these engines assigned a ``malicious'' label to a URL we considered it malicious, and if no engines assigned a ``malicious'' label to a URL we labeled it benign.  We discarded URLs with 1 to 4 anti-virus engine detections.  Our motivation is that URLs may or may not be benign or malicious, and this uncertainty would introduce “label noise” into both the training of our model and our validation of its accuracy. 

Since registry keys and file paths can be ambiguous (e.g., both malware and benignware can write to same path), we took a different approach for labeling file and registry key paths.  First, we labeled our corpus of binaries as either malicious or benign using a voting technique over an ensemble of 60 anti-virus engines, where binaries that had 5 or more anti-virus based detections were labeled as malicious and binaries that had 0 detections were labeled as benign.  We then discarded binaries with between 1 and 4 detections from our dataset, as we regarded the question of whether these binaries were malicious or benign as ambiguous.  Next we inspected the behavioral traces of the resulting 18 million binaries, counting how often each unique file path or registry key was created or written to in both benign and malicious sandbox runs.  Finally, we labeled any file paths or registry keys that only occurred in malicious contexts as malicious, and labeled the other artifacts (benign or partially cases) as benign.

\subsection{Baseline Models}

We implemented two baseline models, one is standard general feature n-gram extractor, where we pool out a set of all possible 1-5 sized n-grams. The second model, used only for URLs, is based on manually extracted features described in \cite{ma2009beyond}. These features include common sense statistics like: URL length, the number of `.' separators in a URL, and categorical lexical features like domain name and URL suffix tokens.  Combined together, they form a very large, but sparse feature vector.

To make training tractable at the scale of millions of examples, we use the feature hashing trick to randomly hash these features into 1024-dimensional vectors.  Our motivation for picking a dimensionality of 1024 was two-fold.  First, in order to tractably train our baseline models on millions of URL examples, a small feature vector is optimal.  Second, given that the output of our novel model’s feature extraction is 1024 dimensional, we were interested in comparing a conventional 1024-dimensional representation of URLs with our deep learning representation, thereby answering the question, which representation is the richest representation for performing malicious URL detection?

The above hashed features are fed directly into a deep MLP model.  This MLP model is identical to our novel neural network model, except that we've stripped off the deep learning feature extraction layers, and replaced the input they provide with our manually constructed 1024-dimensional feature vector.  This design is intended to highlight the potential contribution of our convolutional feature extractor in improving  detection accuracy. 

\subsection{Evaluation}

\begin{figure*}[thpb]
      \centering
      \includegraphics[width=1.0\textwidth]{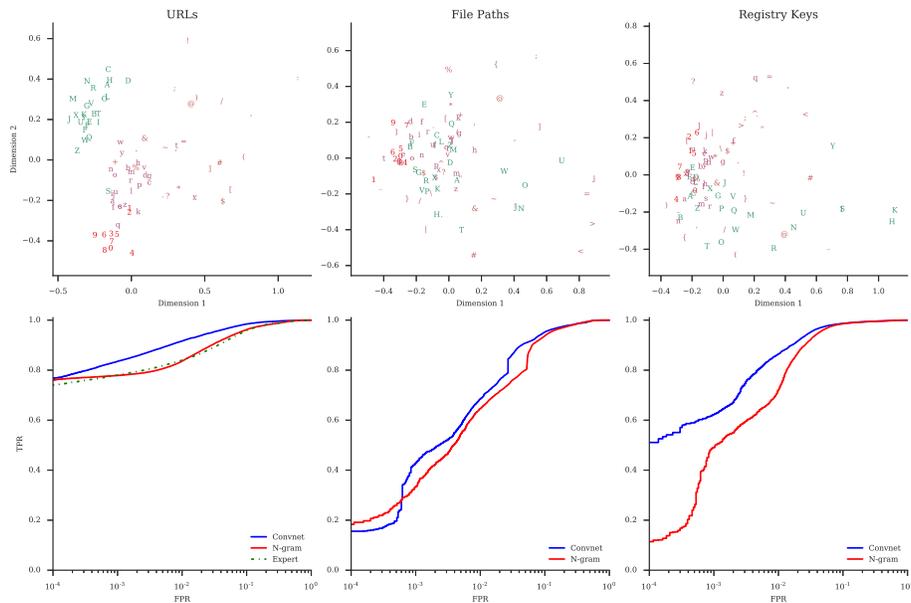}
      \caption{Results for URLs, file paths, and registry keys. Top row shows 2D PCA projection of the normalized character embeddings. Bottom row shows the ROC curve for various models.}
      \label{fig:results}
\end{figure*}

We present our results using a ROC curve between true positive and false positives rates. This measure is independent of the ratio of benign to malware in our dataset, and so is simplest to interpret. We focus on the low FPR rates $10^{-4}$ and $10^{-3}$, which from our experience represent a reasonable deployment threshold.  The ROC curves for all three problems are shown in Fig. \ref{fig:results}, and the specific values are given in Table \ref{tab:results}.

In addition to the ROC curves, we also present the two dimensional PCA projection of the normalized embedding vectors for the all individual characters in Fig. \ref{fig:results}. The capital letters, lowercase letters, numbers tend to cluster together, while important special symbols like ``/'' and ''?'' are fairly separated from the rest of the character. The results support our intuition behind inserting the embedding layer to provide a richer n-gram like detection, by clustering semantically similar characters together.

\begin{table*}[h]
\caption{Validation results}
\label{tab:results}
\begin{center}
\begin{tabular}{|l|l|l|l|l|l|}
\hline
\multicolumn{2}{|c|}{Task} & \multicolumn{3}{|c|}{TPR (at various FPRs)} & AUC \\
\multicolumn{2}{|c|}{ } &  $10^{-4}$ & $10^{-3}$ & $10^{-2}$ &  \\
\hline
\multirow{3}{*}{URLs}
 & Convnet & \textbf{0.77} & \textbf{0.84} & \textbf{0.92} & \textbf{0.993} \\
 & N-gram & 0.76 & 0.78 & 0.84 & 0.985 \\
 & Expert & 0.74 & 0.78 & 0.84 & 0.985 \\
\hline
\multirow{2}{*}{File Paths}
 & Convnet & 0.16 & \textbf{0.43} & \textbf{0.68} & \textbf{0.978} \\
 & N-gram & \textbf{0.18} & 0.33 & 0.65 & 0.972 \\
\hline
\multirow{2}{*}{Registry Keys}
 & Convnet & \textbf{0.51} & \textbf{0.62} & \textbf{0.86} & \textbf{0.992} \\
 & N-gram & 0.11 & 0.49 & 0.72 & 0.988 \\
\hline\end{tabular}
\end{center}
\end{table*}

Our results show that across the board convolution feature extraction outperforms other approaches. For example, at FPR $10^{-3}$ eXpose has $6$\% higher detection rate than n-gram or expert derived features, with even larger improvement on file paths and registry keys problems. The fact that our tuned expert features are not able to outperform n-grams is potentially explain by our large dataset size. This is consistent with the well observed fact in the fields of NLP \cite{shazeer2015sparse} and bioinformatics \cite{ondov2016mash}, that a bag of n-grams is a highly effective representation in itself. The fact that the convolutional network is able to exceed n-gram performance in this large dataset setting suggest that embeddings with convolutional networks can be used as powerful automatic feature extractor.

The overall results for the file paths and registry keys problem are worse than for URLs.  This is not surprising because of the difficulty of properly labeling samples. Recall that our approach was to label any sample that had 0 occurrences in malware data as benign, and everything else was labeled malware. However, estimating if the probability sample being 0 is difficult because typically there is only one observation of the string in the datasets. Furthermore, file paths and registry keys have significantly less training data, which in our experience can significantly decrease our model's generalizability. 

While our CNN model outperform the n-gram model,  one potential explanation is that there are too many collisions in the 1024-sized vector caused by feature hashing. Conversely, the vector is too large for the neural network to capture good relationships between the features. Therefore, we have also done the n-gram experiment with 512 and 2048-sized feature vector. These n-gram experiments yielded worse results than the 1024-sized result, and so are not shown.

We note that, potentially, extensive re-architecture of the n-gram's neural network or switching to a different ML approach could yield better results than we presented. Furthermore, feature importance values, as computed using mutual information between label and individual feature vectors, L1 logistic-regression, random forest, or related methods can be used to significantly reduce the number of n-gram features that are hashed into the vector, thus also potentially improving results. The downside is that these methods in themselves require significant amount of tuning and multiple passes through a very large dataset. The advantage of our end-to-end learning is that we work directly with raw data, and so can simply utilize the same loading of samples in small batches no matter the architecture. With this approach is no \apriori loss of information that is inherent in features engineering, enabling very rapid prototyping.

\section{Conclusions}
\label{sec:conclusions}

We developed and demonstrated the first, to our knowledge, convolutional neural network for extracting automatic features from short string in the context of cybersecurity problems. Using embeddings with convolutions as top layers in our neural network coupled with supervised training, allows us to implicitly extract a feature set that is directly optimized for classification. While similar approaches have been suggested for NLP, eXpose is the first approach that demonstrates how top to bottom deep-learning method can be adapted to several important cybersecurity problems in an adversarial environment, where strings are purposely obfuscated to prevent obvious feature extraction.

One of the major issues during our experimentation was the computational cost of training on longer strings, that prevented us from trying more complex architectures. With current advances in hardware and distributed training modules added to modern frameworks, our results can potentially be further improved with some of the more computationally expensive architecture that we were unable to try.

Looking forward, we hope that ideas integrated into eXpose will help guide the security industry into moving away from expensive feature engineering to directly utilizing already existing labeled datasets for end-to-end learning. As hardware and available datasets improve, the difference between automatically extracted features and traditional feature extraction approaches will only get starker.

\section*{Appendix}

\subsection{Components}

Our convolutional neural network is implemented in Python 2.7 using Keras v1.1 \cite{chollet2015keras}. Below we describe our model's pre-defined set of components (or layers), which are described in terms of Keras build-in layers documented online:

\subsubsection*{$\mathrm{Embedding}(s,m)$}

An embedding layer that takes in a list of $s$ integers representing the URL character list (each unique character is mapped to an associated unique integer in the range $[1,\abs{\Sigma}]$), and outputs a matrix of floating points, where each original scalar integer value is represented by an $m$ dimensional embedding vector. $\abs{\Sigma}$ is the size of the alphabet used to express the URLs. This operation is defined in Keras as $\mathrm{Embedding}(\text{input\_dim=}s)$

\subsubsection*{$\mathrm{Conv}(t, k, n)$}

A filter bank of $t$ $k$-length one dimensional convolution kernels that convolve $n$ adjacent $m$ dimensional vectors, and are immediately followed by a non-linear ReLU activation. Defined in Keras as \newline $\mathrm{Convolution1D}(t, k,\text{input\_shape=}(s,m))$, followed by $\mathrm{Activation}(\mathrm{relu})$. Note, we drop $m$ in our notation, since it can be inferred from the previous layer, or otherwise defined in the text.
\subsubsection*{$\mathrm{BatchNorm}$}

Layer-wise batch normalization. Defined in Keras as \newline $\mathrm{BatchNormalization}(\text{mode=}1)$.

\subsubsection*{$\mathrm{SumPool}$}

Sum of the input along the input length $s$, such that the output size is $k$, given the input size $(s,k)$. The operation is defined in Keras as $\mathrm{Lambda}(\mathrm{f},\text{output\_shape=}(k,))$, where $\mathrm{f}(\textrm{X})=\mathrm{K.sum}(\textrm{X}, \text{axis=}1)$.

\subsubsection*{$\mathrm{Dropout}(p)$}

Dropout with probability $p$ \cite{srivastava2014dropout}. Defined in Keras as $\mathrm{Dropout}(p)$.

\subsubsection*{$\mathrm{Merge}$}

A merge operation that takes output from a previous set of layers, $\{k_1, k_2, k_3, k_4\}$, and concatenates them into a single matrix, $\begin{bmatrix} k_1, k_2, k_3, k_4 \end{bmatrix}^T$. Defined in Keras as $\mathrm{Merge}(\mathrm{...,mode="concat"})$.

\subsubsection*{$\mathrm{Dense}(l)$}

A fully connected linear unit with output size $l$, followed by a ReLU non-linear activation. Defined in Keras as $\mathrm{Dense}(l)$, followed by $\mathrm{Activation}(\mathrm{relu})$.

\subsubsection*{$\mathrm{DenseSigmoid}$}

Last layer used to generate a binary decisions. Same as $\mathtt{Dense}(1)$, but followed by a sigmoid (instead of ReLU) activation, defined in Keras as $\mathrm{Activation}(\mathrm{sigmoid})$.

\section*{Acknowledgment}

We would like to thank Richard Harang and Joe Levy for their valuable feedback on early drafts of the manuscript and Hillary Sanders for in-depth discussion of our URL results.


\bibliographystyle{splncs03}
\bibliography{main}

\end{document}